# Crystal Growth Rates from Molecular Liquids: The Kinetics of Entropy Loss

Richard K. Bowles[1,2] and Peter Harrowell[3*]

[1] *Department of Chemistry, University of Saskatchewan, Saskatoon, SK, Canada S7H 0H1* 3

[2] *Centre for Quantum Topology and its Applications (quanTA), University of Saskatchewan, SK, Canada S7N 5E6* 4

[3] *School of Chemistry, University of Sydney, New South Wales 2006, Australia*

* peter.harrowell@sydney.edu.au

Abstract

It has been established empirically that the rate of addition of molecules to the crystal during crystal growth from the melt is proportional to exp($-|\Delta S_{fus}|/R$) where $\Delta S_{fus}$ is the entropy of fusion. Here we show that this entropic slowdown arises directly from the separation of the entropy loss and energy loss processes associated with the freezing of the liquid. We present a theoretical treatment of the kinetics based on a model flat energy landscape and derive an explicit expression for the coupling magnitude in terms of the crystal-melt interfacial free energy. The implications of our work for nucleation kinetics are also discussed.

## 1. Introduction

There is a puzzle in the literature on the theory of crystal growth from the melt, one related to the kinetic influence of the entropy change on freezing. If we express the steady state speed $V$ of the crystal-liquid interface as the difference between the rate of addition, $k_+$ and the rate of loss $k_-$, we can use the condition of detail balance to eliminate the latter rate and write

$$V = k_+ \left(1 - \exp(\beta \Delta \mu)\right) \tag{1}$$

where $\Delta \mu = \mu_{xtl} - \mu_{liq}$ is the difference in chemical potentials of the crystal and liquid and $\beta = (k_B T)^{-1}$. With Eq.1, the theory of crystal growth is reduced to the specification of the addition rate $k_+$ in terms of molecular parameters. The standard theory of crystal growth from the melt is due to Wilson [1] and, later, Frenkel [2] (WF) who expressed the addition rate per unit area as

$$k_+^{WF} = \frac{Da}{\lambda^2} f \tag{2}$$

where $D$ is the diffusion coefficient, $\lambda$ is the displacement associated with ordering and $a$ is crystal lattice spacing. The fraction $f$ of growth sites was not in the original equation [1] but was later included to account for the influence of surface roughness and intrinsic defects on the mechanism of crystal growth [3]. In the following we shall consider only 'normal' growth of a rough interface and so set $f = 1$. Eq. 2 has been and continues to be widely employed in the literature on crystal growth, e.g. ref. [3-6].

The puzzle arises with the publication in 1988 of a paper by Burke, Broughton and Gilmer (BBG) [7] about a simulation study of the growth kinetics of Lennard-Jones crystals in which they introduced an alternative expression of the addition rate,

$$\begin{aligned} k_+^{BBG} &= \frac{D}{\lambda^2} a \exp\left(-\Delta S_{fus} / R\right) \\ &= k_+^{WF} \exp(-\Delta S_{fus} / R) \end{aligned} \tag{3}$$

where $\Delta S_{fus} / R$ is the molar entropy of fusion (defined as $S_{liq}$-$S_{xtl}$) and $R$, the gas constant.

There is no explanation in ref. 7 for the presence of the extra term involving the entropy change and the single citation associated with this equation is to ref. [8] which makes no mention of entropy of fusion. The puzzle is that this alternative version $k_+^{BBG}$ - with its rather casual introduction, its missing justification and its very much lower usage in the literature relative to $k_+^{WF}$ - is actually the *correct* expression. This was established in an extensive comparison with crystal growth data in 2008 [9]. The success of $k_+^{BBG}$ implies that the widely used rate $k_+^{WF}$ is incorrect. This unresolved situation represents the current state of affairs. The primary goal of this paper is to establish an expression for the rate of addition to the crystal based on an explicit treatment of the coupling between entropy and energy in the phase transformation. A particular focus is to determine under what conditions $k_+^{BBG}$ represents an accurate expression of this addition rate.

In addressing the kinetic consequences of entropy change, we shall explore connections between topics - crystal growth, interfacial free energy and energy landscapes - that are not typically linked. Our description of the organization of the paper, therefore, provides a map of the argument we present. We begin, in Section 2, with a collation of experimental values of $\Delta S_{fus}/R$ for a variety of molecules, and then go on to critically examine the explanations for Eq. 3 that have been published. In this examination we establish the importance of the coupling between entropy and energy change in the transformation between liquid and crystal. In Section 3 we argue that a quantitative measure of this coupling is implicit in the existing data for the crystal-liquid interfacial free energy. In Section 4 we demonstrate that the kinetic consequences of entropy-energy coupling can be expressed in the energy landscape picture of complex kinetics as developed for protein folding. Extending this idea, we present in the following Section an explicit solution for the kinetics of ordering when the entropy is not strongly coupled to the energy change and, in doing so, demonstrate how the $\exp(-\Delta S_{fus}/R)$ term appears quite naturally in the kinetics associated with entropy loss. In Section 6 we consider types of growth kinetics with stronger coupling between energy loss and entropy loss and the enhanced kinetics that results and, in Section 7, we consider how to select the degrees of freedom associated with crystal growth.

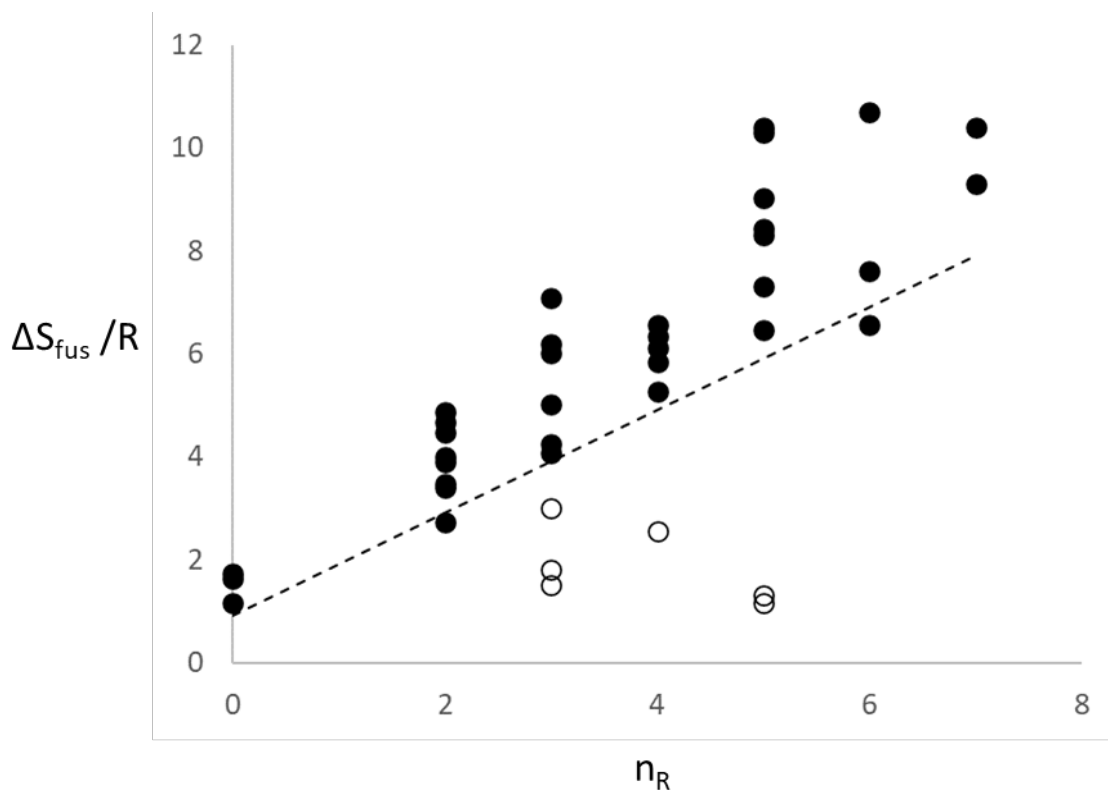

**Figure 1.** The value of the entropy of fusion, $\Delta S_{fus}/R$, (filled circles) for molecules as a function of $n_R$, the number of rotational degrees of freedom. Note that $n_R$=2 for diatomics and rigid rods, $n_R$=3 for rigid non-linear polyatomics and additional rotations arise from internal rotations about single bonds. We do not count those rotations whose only consequence is to rearrange hydrogen atoms. The dashed line represents an increase of 1.0 for each rotation. The values of $\Delta S_{fus}/R$ for plastic crystals are indicated by open circles. The data used in this plot are provided in Table A1 in the Appendix.

## 2. The Entropy of Fusion and the Rate of Crystal Addition

Whether or not the difference between $k_+^{WF}$ and $k_+^{BBG}$ matters depends on the magnitude of the entropy of fusion. Experimental values of $\Delta S_{fus}/R$ for a range of molecules are presented in Fig.1 (and tabulated in Table 1 in the Appendix) and organised in terms of the number of rotational degrees of freedom $n_R$. We count the rotations as follows. First, we consider the rigid body rotations of the molecule (2 for a linear molecule and 3 for a nonlinear molecule). To this we add the internal rotation around single bonds, excluding those that only reorganise hydrogen atoms e.g. the rotation of a methyl group. For atomic liquids (i.e. $n_R = 0$) we find $1 \leq \Delta S_{fus}/R \leq 2$. This range of values is consistent with the estimate by Hirschfelder, Stevenson and Eyring [10] who considered a simple model of melting in which, below the melting point, each particle is localised in one of $N$ cells that fill the volume while above $T_m$ each particle can sample the entire volume. The value of this communal entropy for a mole of particles is $R$. The observed value of $\Delta S_{fus}/R$ is expected to exceed the value of this communal entropy due to the expansion that accompanies melting at constant pressure. Pure metals exhibit an approximately constant value of $\Delta S_{fus}/R \sim 1.16$, an observation that is referred to as Richard's rule [11]. Two model atomic liquids, hard spheres and Lennard-Jones particles, also lie within the expected range with values of $\Delta S_{fus}/R$ of 1.64 [12] and 1.7 [13], respectively. For $1 \leq \Delta S_{fus}/R \leq 2$, $k_+^{BBG}$ is smaller than $k_+^{WF}$ by a factor of 2-7, a difference that is small enough that it can be lost in the uncertainty inherent in the simple theoretical treatment of growth and the uncertainty of the experimental data. It is interesting to note here that phase change materials, characterised by their rapid crystal growth [14], also appear to have small entropies of fusion, despite the complexity of their composition. Two examples: $Ge_2Sb_2Te_5$ has a $\Delta S_{fus}/R = 1.62$ [15] and $AgInSbTe$ has a $\Delta S_{fus}/R = 2.48$ [16].

6The situation changes dramatically when we turn to the growth rates of molecular crystals into their melts. As shown in Fig. 1, $\Delta S_{fus}/R$ increases significantly with the number of rotations $n_R$. The minimum increases in $\Delta S_{fus}/R$ per rotation is ~ 1 as represented by the dashed line in Fig. 1. This minimum line corresponds to the limit in which freezing only imposes the minimal constraints necessary to obstruct the $n_R$ rotations. Entropy loss greater than this minimum might arise through large density change, additional constraints on molecular vibrations, etc. The large entropy change leads to substantial differences in the magnitudes of $k_+^{BBG}$ and $k_+^{WF}$. Glycerol, for example, with $n_R = 6$ (i.e. 3 rigid body rotations and 3 internal rotations about carbon single bonds) has an entropy of fusion of $\Delta S_{fus}/R = 7.6$ so that $\frac{k_+^{BBG}}{k_+^{WF}} \sim 5 \times 10^{-4}$. Indomethacin with $n_R = 5$ freezes into the $\alpha$ crystal phases with $\Delta S_{fus}/R = 10.3$ and, thus, $\frac{k_+^{BBG}}{k_+^{WF}} \sim 3 \times 10^{-5}$.

The increase in the entropy of fusion with $n_R$ depends on these rotations being constrained in the crystal. In plastic crystals [17,18], the molecules retain considerable disorder in the rotational degrees of freedom and, hence, the entropy cost of freezing is significantly reduced. In Fig. 1 we have included data for a number of molecules that form plastic crystals (see Table A1 in the Appendix) demonstrating their anomalously small values of $\Delta S_{fus}/R$ relative to the analogous molecule with an ordered crystal.

In 2008, Ediger, Harrowell and Yu [9] examined the correlation between a reduced attachment rate and $\Delta S_{fus}/R$. The reduced rate is defined as

$$k_+^{red} = \frac{V\tau_\alpha}{a(1-\exp(\beta\Delta\mu))}$$
$$= \frac{k_+}{k_+^{WF}} \qquad (4)$$

where V is the measured growth rate, $\tau_\alpha$ is the stress relaxation time ($\sim \eta / E$ where $\eta$ is the shear viscosity and E is the Youngs modulus) and *a,* the crystal lattice spacing. If the attachment rate was well described by $k_+^{WF}$, then $k_+^{red} \approx 1$, independent of the choice of molecule. In Fig. 2 we have replotted the data from ref.9. The data strongly supports the $k_+^{BBG}$ prediction, i.e.

$$k_+^{red} = \exp\left(-\Delta S_{fus} / R\right) \qquad (5)$$

for a range of systems that include organic molecules and inorganic materials (including oxides like $SiO_2$ and $GeO_2$). A telling confirmation of Eq. 5 is the data for succinonitrile. This organic molecule is of similar size to a number of the other organic compounds included but exhibits an addition rate ~ 3 orders of magnitude greater the other organics. This difference is fully accounted for by $k_+^{BBG}$ and the reduced entropy of fusion associated with its plastic crystal phase (see Fig. 1). There are exceptions to Eq. 5 - the complex inorganic compounds: anorthite ($CaO \cdot Al_2O \cdot 2SiO_2$), cordierite ($5SiO_2 \cdot 2Al_2O \cdot 2MgO$) and diopside ($CaO \cdot MgO \cdot 2SiO_2$) exhibit faster rates of attachment than predicted by $k_+^{BBG}$.

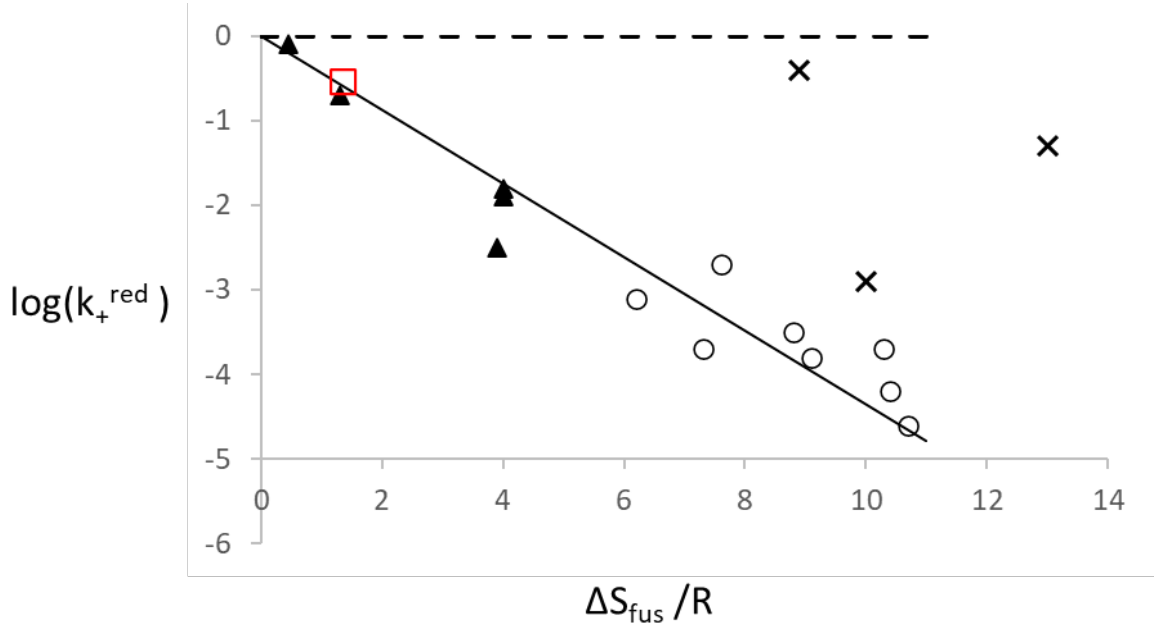

**Figure 2.** The dependence of $\log\left(k_+^{red}\right)$ on $\Delta S_{fus}/R$ from experimental data as cited in ref. 9. The figure includes data for organic molecules (open circles) and inorganic compounds (filled triangles). The solid line corresponds to $\log\left[\exp\left(-\Delta S_{fus}/R\right)\right]$ and the dashed horizontal line corresponds to the WF result. Three outlying results: anorthite, cordierite and diopside, are indicted by crosses. The data for succinonitrile [19], a plastic crystal, is included as a red open square.

With the validity of $k_+^{BBG}$ empirically established, we shall now consider the two justifications of the expression that we have found in the literature. The first is that of the original authors. In 1995, Jackson, Gilmer and Temkin [20] described a Monte Carlo treatment of crystal growth that begins with an explicit expression for the rate of loss from a crystal surface in the form

$$k_-^{BBG} = \nu \exp(-\beta \Delta E) \qquad (5)$$

where $v$ is a frequency that, here, we set to $v = \dfrac{Da}{\lambda^2}$ and $\Delta E$ is the energy increase required to move a molecule from the crystal to the liquid, a quantity that can generally be equated with the molecular heat of fusion. Eq.5 is a strong assumption based on the sharp interface model, due to Kossel [21] and Stranski [22], for crystal growth from the vapor or solution rather than the diffuse interface [23] between the crystal and its melt. It contains an implicit picture of a jump from a surface bound particle to one completely detached from the crystal. Having made this this assumption, the form of $k_+^{BBG}$ in Eq. 3 follows directly as the simple consequence of imposing detailed balance. The effect of Eq. 5 is to divide thermodynamic control of growth kinetics into a loss governed by the enthalpy difference and addition which, by default, is controlled solely by the entropy loss. We shall return to this question of the separation of entropic and energetic control in the following Section.

An alternative justification of Eq. 3 has been provided by Kashchiev in his book on nucleation theory [24]. His argument goes as follows. We start with the rate of addition to a cluster of size $n$, $k_{+,n}^K$, written as

$$k_{+,n}^K = \gamma_n \left( D Z_1 / d_o \right) A_n \tag{6}$$

where $\gamma_n$ is a 'sticking' coefficient, $D$ is the diffusion coefficient, $A_n$ is the cluster surface area and $Z_1$ is the concentration of crystalline clusters consisting a single molecule, i.e. a molecule with a crystal-like arrangement of neighbours who, themselves, are not in crystalline environments. Eq. 6 asserts that 'regular' molecules from the bulk liquid do not attach to the growing crystal surface, only those that are in a crystal-like environment prior to attachment. This is an essential feature of this approach.

At equilibrium

$$Z_n = \frac{1}{v_o}\exp(-\beta W(n)) \qquad (7)$$

where $v_o$ is the volume per molecule in the liquid and W(n) is the work required to form the n-cluster, W(n), is written as

$$W(n) = n\Delta\mu + \phi(n) \qquad (8)$$

where $\phi(n)$ is the interfacial free energy of the n-cluster. Kashchiev [24] then uses the following expression for the surface contribution,

$$\phi(n) = n\varepsilon - E_n \qquad (9)$$

where $\varepsilon$, the energy difference between a molecule in the liquid and crystal phases, is essentially the molecular heat of fusion and $E_n$ is the energy released by the disassociation of the n-cluster into separate molecules in the liquid phase. The idea is that this difference represents the surface contribution (included in $E_n$ but not $n\varepsilon$) is treated purely energetically. (It clearly cannot be applied, for example, to a system of hard particles.) Next, the energy to transform the local coordination of a molecule into a crystal-like one is set to zero, i.e. *$E_1$ = 0*. It is not clear how this is justified but, with this choice, Kashchiev assumes that the entropy loss occurs with no change in energy, implicitly decoupling entropy loss from energy loss. With *$E_1$=0*, the resulting expression for *W(1)* is obtained from substitution Eq. 9 into Eq. 8 and setting *n = 1* to give

$$W(1) = \Delta\mu - \varepsilon = T\Delta s_{fus} \qquad (10)$$

(since $\Delta\mu = \varepsilon - T\Delta s_{fus}$) so that the work to create the isolated crystal-like molecule is $Ts_{fus}$ (where $\Delta s_{fus} = \Delta S_{fus}/N_A$) which, when substituted in Eqs. 6 and 7, gives us the BBG result. The alternative would be to assume that any molecule can be added, in which case

$$Z_1 = 1/v_o, \quad (11)$$

a choice that would result in the $k_+^{WF}$ expression. A particular feature of the derivation of $k_+^{BBG}$ in ref. 24 is that the kinetics of entropy loss is 'invisible' since it occurs prior to the start of the explicit kinetic scheme involving the addition of crystal-like molecules to the growing crystal.

## 3. The Coupling of Entropy and Energy Change in the Crystal-Liquid Interfacial Energy

The separation of entropy loss from energy loss that appears implicitly in both the Jackson et al [20] and the Kashchiev [24] treatments of the $k_+^{BBG}$ expression highlights the importance of the coupling of entropy and energy changes of the kinetics of ordering processes. This coupling is encoded in the topography of the energy surface, either potential energy or free energy, over some high dimensional space of particle configurations [25,26]. As such, the coupling can be manifest in static properties as well as dynamic ones. In this Section we shall consider two important applications involving entropy-energy coupling: the crystal-melt interfacial energy and the kinetics of protein folding.

Given the small density difference between the crystal and melt in pure metals, the large magnitudes of their crystal-melt interfacial free energy, established by the nucleation studies of Turnbull and Cech in 1950 [27], came as a surprise [28]. To account for this interfacial free energy, Spaepen and Meyer [29, 30] considered a structural model of the crystal-melt interface to assess how ordering (i.e. entropy loss) and energy change were related through the interface. Here we shall present a simple model of the argument to establish the connection between this coupling and the interfacial free energy. Pirzadeh et al [31] have presented a similar perspective.

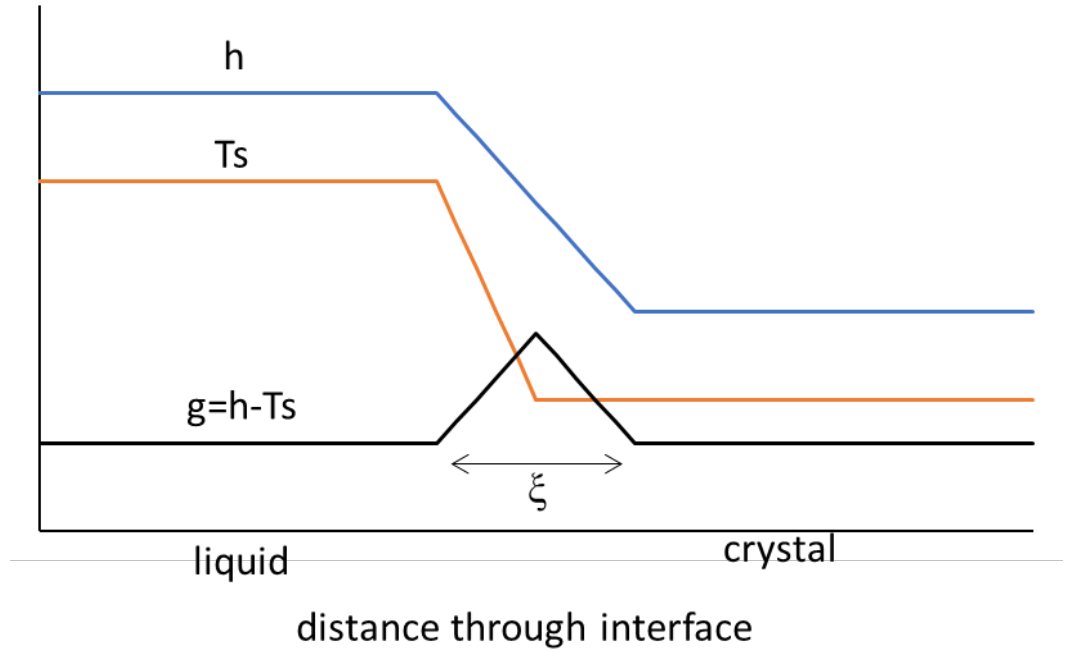

**Figure 3.** Schematic plot of the enthalpy density h and entropy density s (presented here as the product Ts) profiles as function of the normal displacement through a planar crystal-melt interface at coexistence (i.e. $g_{xtl} = g_{liq}$). Also plotted is the profile of the free energy density g = h-Ts. The interfacial width $\xi$ is indicated.

Assuming a local equilibrium model [32], we can imagine the profiles of the enthalpy density *h* and the entropy density *s* through a planar interface as sketched in Fig. 3 where $-m_h$ and $-m_{Ts}$ are the gradients of h and Ts, respectively, through the interface. The ratio $c = \dfrac{m_h}{m_{Ts}}$ establishes the degree to which entropy and enthalpy are coupled through the interface with *c* =*1* being perfectly coupled and *c* ~ *0* being completely uncoupled. Even though both quantities decrease monotonically going from liquid to crystal, the local free energy density, $g = h-Ts$, exhibits a peak in the interface region as a consequence of the entropy decreasing more steeply than the enthalpy, i.e. as a consequence of the coupling between enthalpy and entropy being less than perfect as measured by $c < 1$. In this simple model, we can express analytically the dependence of the surface free energy $\gamma$ on the coupling $c$. On integrating $g - g_{bulk}$ through the interface we find the free energy of the interface $\gamma$ to be

$$\begin{aligned}\gamma &= \Delta h_{fus} \frac{\xi}{2}(1-c) \\ &= \frac{\Delta H_{fus}}{N_A} \frac{\xi}{v_o} \frac{1-c}{2}\end{aligned} \qquad (12)$$

where we have used $m_H = \Delta h_{fus}/\xi$ with $\xi$ being the interfacial width as indicated in Fig. 3. The interfacial free energy would vanish in the case of perfect coupling (i.e. $c =1$) and it is a maximum in the completely decoupled case of $c$ approaching zero. This latter case corresponds to the situation in which the loss of entropy due to structural ordering occurs completely before any loss of energy is achieved.

We can compare Eq. 12 with the empirical expressions for the crystal melt interfaces of atomic liquids due to Turnbull [33] and Laird [34], i.e.

$$\gamma = \frac{\Delta H_{fus}}{N_A} \frac{1}{v_o^{2/3}} C_T \qquad \text{Turnbull [33]} \qquad (13)$$

where $C_T = 0.45$ for close packed metals, and

$$\begin{aligned}\gamma &= T_m \frac{\Delta S_{fus}}{N_A} \frac{1}{v_o^{2/3}} \frac{1}{2} \\ &\approx T_m \frac{R}{N_A} \frac{1}{v_o^{2/3}} \frac{1}{2}\end{aligned} \qquad \text{Laird [334]} \qquad (14)$$

where we have arrived at Laird's expression by using the estimate for atomic liquids [12] of $\Delta S_{fus} \approx R$. On equating the empirical equation, Eq. 13, with the result Eq. 12 from our model we find

$$\frac{\xi}{v_o} \frac{1-c}{2} = \frac{C_T}{v_o^{2/3}} \tag{15}$$

which, if we assume $\xi \sim v_o^{1/3}$, offers a potentially useful estimate of the entropy-energy coupling $c$, i.e.

$$\begin{aligned} c &= 1 - 2C_T \\ &= 1 - \frac{2\gamma v_o^{2/3} N_A}{\Delta H_{fus}} \end{aligned} \tag{16}$$

Based in Eq. 16, $c$, the entropy-enthalpy coupling, for atomic liquids lies in the range $c \sim 0.1$. This evidence of weak coupling is consistent with the weak coupling implicitly assumed in the kinetic theories of Jackson et al [20] and Kashchiev [24]. Turnbull [33] reported a smaller value of $C_T$ ($\sim 0.32$) for nonmetals such as Ge and Bi. Feng and Laird [35] have reported a similar value of $C_T$ for succinonitrile. Again using Eq. 16, these results indicate $c \sim 0.36$, a greater entropy-energy coupling than that of the closed packed metals.

## 4. The Coupling of Entropy and Energy in Energy Landscapes

The question of the kinetic consequence of the coupling between entropy and energy change has perhaps been most intensely studied in the context of the kinetics of protein folding. Where our discussion has kept returning to the presence of weak coupling governing the attachment kinetics in crystal growth, the story of protein folding kinetics is essentially one of emphasising the necessity of significant entropy-energy coupling. The problem posed by weak coupling in the context of protein folding was pointed out by Levinthal in 1969 [36] when he noted that, *if* the search for the native folded state proceeds via random sampling of the configuration space, then the time required to fold a polypeptide of even modest length would exceed $10^6$ years, a far cry from the observed folding times typically measured in seconds or minutes.

The Levinthal scenario corresponds to a model landscape that is flat save for the target structure which sits at the bottom of a localised 'hole'. Such a landscape is referred to as a 'golf course' landscape [37]. Real landscapes for globular proteins are expected [38] to be i) 'rough', i.e. exhibiting multiple additional metastable minima, and ii) 'funnelled' so that these minima are organised in descending energy around the target structure (see Fig. 4). In the funnel landscape, the overall energy gradient can 'steer' the structural search towards the global minimum and, thus, significantly reduce the time scale for ordering. These two types of landscapes – golf course and funnel – provide graphical representations of weak and strong coupling between entropy change and energy change as shown in Fig. 4. In the following two Sections we shall consider the application of the golf course and funnel landscapes to molecular crystal growth.

We must clarify whether we are using a potential energy or a free energy landscape to describe crystallization. Wales and Bogdan [39] have provided a useful discussion on the difference. The potential energy landscape is the natural landscape to describe kinetics when the initial and final states are described by single minima. Chemical reactions and defect motion in a crystal are examples of such processes. Order-disorder transitions, in contrast, involve at least one state characterised by averages over many local potential energy minima and so need to be treated with a free energy landscape. The choice for crystal growth from the melt, we suggest, is not quite so clear cut. We could, for example, consider the crystal growth rate in terms of an individual local minimum (i.e. inherent structure) of the liquid adjacent to the interface. This approach is the basis of a recent treatment [40,41] of the ultrafast crystal growth observed in pure metals and molten salts. In this treatment, it is the potential energy landscape and the degree of order present in the interfacial inherent structures that are essential to explain the anomalously high growth rate.

A more generic issue arises when we select which degrees of freedom we wish to treat via explicit dynamics. We should consider all of the degrees of freedom that are associated with the ordering of the liquid. These can be identified as those degrees of freedom which suffer a significant increase in constraint during crystallization. Omission of any of these degrees of freedom from the dynamics is justified only if their relaxation dynamics is much faster than the others that they can be treated as continuously equilibrated. In extracting these degrees of freedom from those that are averaged over to generate the free energy surface, we effectively strip the entropic contribution from free energy, leaving us with a surface that is, in effect, the potential energy, averaged over the fluctuations of the 'uninteresting' degrees of freedom (i.e. those unaffected by the transition from liquid to crystal). So, in summary, the choice – potential energy or free energy surface – becomes moot once we elect to retain those degrees of freedom directly associated with the entropy of fusion as explicit dynamic variables since the resulting constrained free energy is essentially just the potential energy averaged over the fast uninteresting degrees of freedom.

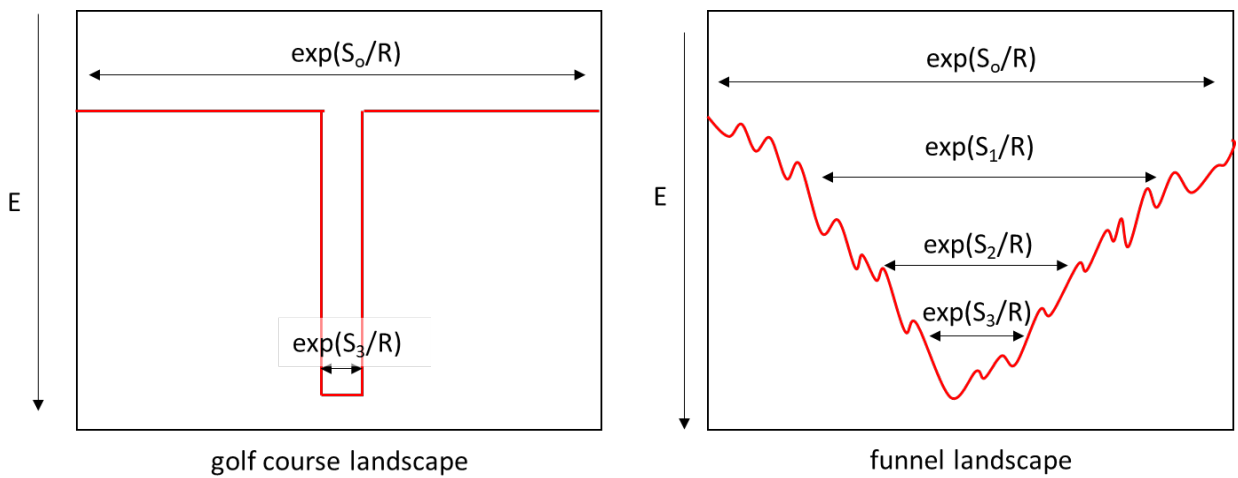

**Figure 4.** Schematic representations of the golf course and funnel landscapes demonstrating the weak and strong coupling, respectively, between the loss of energy and the loss of entropy.

## 5. The Rate of Addition in the Absence of Entropy-Energy Coupling: Freezing in a Golf Course Landscape

The golf course landscape provides a useful starting point for the development of a theory of the rate of crystal addition; it is both simple and, with its weak entropy-energy coupling, consistent with our treatments of the crystal-melt transition. Here we consider a solvable version analysed by Bicout and Szabo [42].

If we neglect the correlations between molecules, then each molecule must find its target state by a search of its own configuration space, i.e. its translations, rigid body rotations and internal rotations, a total of $d$ degrees of freedom. Let this single molecule configuration space be represented then as a d-dimension sphere of radius $R_d$ with reflecting boundary. (The assumption of a spherical configurations space is introduced to simplify the analysis. A more realistic anisotropic phase space will quantitatively change the rates but not alter the dependence on the entropy change.) At its center lies the ordered state consisting of configurations lying within a distance r of the origin. If we write $a = r/R_d$ then the entropy loss on ordering is

$$\Delta S_{fus} = -R \ln(a^d) \tag{16}$$

Note that Eq. 16 represents an explicit statement of the perspective, explained in the previous Section, that the degrees of freedom retained as dynamic degrees of freedom are exactly those whose constraint in the ordered phase gives rise to the entropy of fusion. Following ref. [42], we model the particle's dynamics through its configuration space as a random walk governed by an isotropic diffusion coefficient D. (We shall return to this assumption in Section 6.) The rate of ordering corresponds to the net flux of probability into the ordered configuration space. We shall first consider the simpler case where ordering is irreversible, i.e. the 'black hole' golf course. In this case, the inner surface of radius r is treated as an absorbing boundary. If $a << 1$, Bicout and Szabo [42] found the rate of ordering for the golf course (GC) landscape to be

$$\begin{aligned} k_+^{GC} &= d(d-2)Da^{d-1}/aR_d^2 \\ &= d(d-2)\frac{D}{r^2}\exp(-\Delta S_{fus}/R) \end{aligned} \tag{17}$$

The proportionality of the rate with respect to $a^{d-1}$ (and, hence, the reason for the presence of $\Delta S_{fus}$) can be understood by analogy with rate of escape of a diffusing particle through a narrow opening. The more restricted the extent of the transition state along direction orthogonal to the reaction coordinate relative to the analogous extent in the initial state, the lower the probability of a stochastic trajectory achieving this transition point. This aspect of kinetics has been referred to as an 'entropy' barrier [42,43]. While formally correct, this terminology can obscure the fact that $\Delta S$ can be altered by correlations and/or constraints in the initial state configuration space, features well removed from the transition state itself [44].

To connect Eq. 17 with the previous expressions for the addition rate we multiply $k_+^{GC}$ by the lattice spacing *a* to make it a velocity and then equate r ~ λ. The reasoning here is that r corresponds to magnitude of fluctuations retained in the ordered phase and so represents a reasonable estimate of, at least, the lower bound of the displacement λ associated with ordering. With this choice, we find

$$ak_+^{GC} = d(d-2)k_+^{BBG} \qquad (18)$$

Comparing this approach to the previous justifications for $k_+^{BBG}$, we have invoked a separation of entropy and energy, analogous to that implied in ref. 23 but here it is explicit in the choice of landscape. Where the Kashchiev [24] treatment provides no kinetic model of how the entropy of a molecule is lost prior to addition, our treatment above treats this loss explicitly and relates the kinetics to a well-defined approximate energy landscape.

The inclusion of the reverse process on the golf course landscape does not change things too much. For a reversible process with a target state energy -δ with *d* = 3 [42], the equilibrium constant can be expressed in terms of the fraction *p* in the target state as

$$K_{eq} = \frac{p}{1-p} = \frac{a^3}{1-a^3} e^{\beta\delta} \qquad (19)$$

The diffusion dynamics can be approximated by a two-state model – ordered and disordered – with transition rates $k_+$ and $k_-$ where

$$k_+^{GC} = \frac{D}{R_d^2}\left(\frac{a^2}{15K_{eq}} + \frac{(1-a)^2(5+6a+3a^2+a^3)}{15a(1+a+a^2)}\right)^{-1} \qquad (20)$$

and $k_- = \dfrac{k_+}{K_{eq}}$. In the limit of small *a,* we find

$$k_+^{GC} = 3\frac{Da^3}{r^2}\left(\frac{1}{1+e^{-\beta\delta}/5}\right) \quad (21)$$
$$\approx 3\frac{Da^3}{r^2}$$

and

$$k_-^{GC} = 3\frac{D}{r^2}\exp(-\beta\delta) \quad (22)$$

Since $e^{-\beta\delta}$ lies between 0 and 1, the value of $k_+$ varies little from the irreversible case Eq. 17. Generalising to d degrees of freedom we have

$$k_+^{GC} = d(d-2)\frac{D}{r^2}\exp(-\Delta S_{fus}/R)$$
$$k_-^{GC} = d(d-2)\frac{D}{r^2}\exp(\Delta H_{fus}/RT) \quad (23)$$

With this derivation of $k_+^{GC}$ and $k_-^{GC}$ we have arrived at picture in which addition is governed by entropy loss while detachment is governed by energy increase, exactly as assumed by Jackson et al [20], but without having had to rely on any assumption about the physical details of the interface or the explicit mechanism of particle loss from the surface.

## 6. Anomalously Rapid Crystal Growth and the Role of Funnel Landscapes.

In the literature on protein folding, the golf course model is introduced as a cautionary tale [45] – a demonstration of the unphysically slow kinetics in the absence of any energetic guidance of the entropy loss – and, hence, to be dismissed in favor of some sort of funnelled landscape. In the case of crystal growth, the relative merits of the two types of landscape are inverted, with the weak coupling limit, as expressed by the golf course landscape, proving to be a good account of the ordering process. That said, there are examples of crystal growth that exhibit positive deviations from $k_+^{BBG}$ (or $k_+^{GC}$). Our goal in this Section is to consider whether these deviations indicate the presence of a funnel landscape for crystal growth

Since the primary impact of the funnel landscape is to restrict the size of the configuration space a system must explore to arrive at the ordered state, it follows that $k_+^{WF}$ represents a rough upper bound of the addition rate achievable by this reduction of the search space. By this reasoning, the criterion that a system must meet to be a potential candidate for funnel-mediated crystal growth is that $k_+^{BBG} < k_+ < k_+^{WF}$, i.e. $k_+^{red}$ lies between the two lines in Fig. 2. We note that the complex oxides anorthite and cordierite meet this condition. Such landscapes might be associated with preorganization of local structure in the liquid adjacent to the growing interface or the influence of long-range electrostatic interactions and so represent interesting subjects for further study.

The two most notable examples of anomalously fast crystal growth – metals and simple molten salts– exhibit addition rates that come close to or exceed $k_+^{WF}$. This enhanced crystal growth kinetics is also characterised by a broad peak in the growth rate V as a function of temperature [46]. These features are the signature of growth that is not controlled by activated processes. This absence of diffusion control was noted for FCC crystal growth for a Lennard-Jones system by Broughton, Gilmer and Jackson [8] in 1982. In the language of landscapes, this indicates not only the presence of a potential energy funnel but that this funnel is 'smooth', i.e. lacking the energy barriers to reorganization that characterizes the 'rough' funnels [47] in protein folding. The absence of barriers was explicitly demonstrated in the case of freezing of metals into FCC crystals and molten salts into the NaCl structure where the crystal front was found to advance, in simulations, during energy minimization [40,41].

We conclude this discussion with two observations: i) Energy landscape funnels clearly play a central role in the rapid crystallization of atomic liquids and, probably, phase change materials. These funnels, however, differ from those discussed in the protein folding literature in that they are generally smooth so that the crystal growth is largely decoupled from activated kinetics. ii) While we are unaware of any examples of crystallization kinetic corresponding to a rough funnel landscape, it is an interesting open question with potential candidates to be found in the complex oxides.

## 7. On Selecting Degrees of Freedom

In the golf course landscape model of crystal growth that we have presented, the entropic barrier to molecular addition at the crystal interface depends on the selection of the degrees of freedom that are to be treated as dynamic variables. Given the importance of this selection, what happens if we get it wrong, i.e. omit a relevant degree of freedom or include a redundant one? For each degree of freedom $\alpha$ there is a quantity $a_\alpha = r_\alpha / R_d$ corresponding to the reduction of configuration space associated with that degree of freedom due to ordering. These terms are related to $\Delta S_{fus}$ by

$$\Delta S_{fus} = R \ln\left(\prod_\alpha^d a_\alpha\right) \tag{24}$$

A degree of freedom is counted as relevant if $a_\alpha \ll 1$. Note that redundant degrees of freedom (i.e. $a_\alpha \sim 1$) make no contribution to the entropy change. The omission of a relevant degree of freedom will be indicated by the failure of the equality in Eq. 24. One way of avoiding this problem would be to forego selection entirely and include all degrees of freedom. This would automatically see Eq. 24 satisfied but what about the kinetics? Our expression, Eq. 17, for $k_+^{GC}$ made use of the $a \ll 1$ limit, a limit that does not apply to redundant degrees of freedom. The general expression for the ordering rate is provided in ref. 42 from which we can extract the following approximate expression $(k_+^{GC})_\alpha$ for the ordering rate for the $\alpha$ degree of freedom that is valid in both limits of $a$,

$$(k_+^{GC})_\alpha \propto \frac{D}{r^2} \frac{a_\alpha}{(1-a_\alpha)^2} \tag{25}$$

which diverges $a_\alpha \to 1$. This means that irrelevant degrees of freedom will relax very quickly and so will not influence the slow processes associated with ordering. While formally elegant, the treatment of a large number of rapidly relaxing redundant degrees of freedom would considerably complicate the theory and risk obscuring the insight to be gained from the treatment of the smaller number of physically relevant quantities.

## 8. Discussion and Conclusion

In this paper we have presented evidence from both empirical growth rates and the measured interfacial free energies that molecular crystallization is characterized by a weak coupling between entropy loss and energy loss. In physical terms, we might think of this as two-stage ordering, a process in which the liquid first organizes itself into a loosely ordered arrangement, sufficiently constrained to result in a significant decrease in entropy but still at the liquid density, and then this structure is compacted to the final crystal density, a process characterised by a significant decrease in the energy. Specific microscopic mechanisms aside, the decoupling of ordering and energy is, we have argued, quite generic in molecular liquids and gives rise to a slower addition rate, relative to the standard model, because of the need to locate this loosely ordered configuration by a random search without the benefit of a guiding potential surface. This time required for the random search increases exponentially with the number of molecular rotations – both rigid body and internal – that are constrained by crystallization. This result resolves the puzzle of the enormous difference between the growth rates of atomic and molecular crystals [46], a difference of up to several orders of magnitude that could not be explained by either differences in diffusion constants or thermodynamic driving force.

We have established that a very simple model of the energy landscape – a flat surface punctured by a narrow hole corresponding to the ordered state - provides a reasonable representation of molecular crystal growth. This model is capable of being extended to include the quantitative differences between the dynamic degrees of freedom, both in restraint (the quantities $r_\alpha$) and intrinsic rates, $\nu_\alpha$ in a transparent and systematic fashion. These extensions would require the reanalysis of the d-dimension diffusion problem [42] to include deviations from isotropy of the boundary conditions and the diffusion constant D, respectively. This we leave to future work.

While we have addressed crystal growth in his paper, it is important to note that the validation of $k_+^{BBG}$ as the rate of addition has an immediate impact on crystal nucleation. Recently [44], we demonstrated that the inclusion of variations in the degree of order in pre-critical crystal clusters reduced the nucleation rate. Here we present a more general statement of that result. It is straightforward to show [24] that the stationary distribution of crystal clusters is unaltered if we multiply every rate – both addition and loss - by a factor $\exp(-\Delta S_{fus}/R)$, while the steady state nucleation rate is decreased by this factor. It follows that the use of $k_+^{BBG}$ for the addition rate in the kinetics of crystal clusters will result in actual nucleation rate I being reduced, relative to the value obtained using $k_+^{WF}$, i.e.

$$I = \exp\left(-\Delta S_{fus}/R\right) I^{WF}. \tag{28}$$

As the new factor can reduce the rate of nucleation in molecular liquids by up to 5 orders of magnitude, it is a significant slowing down of crystal nucleation by molecules, a result of particular relevance to the glass forming ability of these liquids [52].

In conclusion, we have presented a general and transparent treatment of crystal growth kinetics of molecular liquids based on the separation of entropy loss and energy loss in the growth process. It is our hope this new formalism will prove valuable for exploring the role of molecular shape and flexibility in determining the kinetics of crystallization.

**Acknowledgements**

In a world awash with Lennard-Jonesium, it is a pleasure to acknowledge the eloquent case Pablo Debenedetti has made for the richness and subtleties of molecular liquids and their transitions. RKB acknowledges NSERC grant RGPIN-2019-03970 for financial support.

**Appendix**

**A1. Data for Figure 1.**

| Material | $\Delta S_{fus}/R$ | $n_R$ | refs |
|---|---|---|---|
| **Ordered Crystals** | | | |
| pure metals (avge) | 1.16 | 0 | 11 |
| Lennard-Jones | 1.74 | 0 | 13 |
| hard spheres | 1.64 | 0 | 12 |
| CO | 2.72 | 2 | 49 |
| $O_2$ | 3.47 | 2 | 49 |
| $F_2$ | 3.41 | 2 | 49 |
| $Cl_2$ | 4.47 | 2 | 49 |
| $Br_2$ | 4.86 | 2 | 49 |
| HCN | 3.89 | 2 | 49 |
| FCN | 4.00 | 2 | 49 |
| $CO_2$ | 4.66 | 2 | 49 |
| furon | 4.08 | 3 | 49 |
| thiophene | 5.01 | 3 | 49 |
| benzene | 4.25 | 3 | 49 |
| azulene | 6.18 | 3 | 49 |
| phenanthrene | 6.02 | 3 | 49 |
| anthracene | 7.09 | 3 | 49 |
| biphenyl | 6.55 | 4 | 49 |
| 1,3-butadiene | 5.85 | 4 | 49 |
| 1,2-butadiene | 6.12 | 4 | 49 |
| 1-butene | 5.27 | 4 | 49 |
| tetralin | 6.33 | 4 | 49 |
| indomethacin α | 10.3 | 5 | 50 |
| indomethacin γ | 10.4 | 5 | 50 |
| indomethacin δ | 7.3 | 5 | 50 |
| 1,1-diphenylethane | 8.3 | 5 | 49 |
| 1,2-diphenylethane | 9.02 | 5 | 49 |
| n-butane | 6.47 | 5 | 49 |
| tris(naphthylbenzene) | 10.7 | 6 | 51 |
| glycerol | 7.6 | 6 | 52 |
| triethylbenzene | 6.65 | 6 | 53 |
| 1-octene | 10.74 | 7 | 54 |
| n-heptane | 9.3 | 7 | 55 |
| **Plastic Crystals** | | | |
| camphor | 1.82 | 3 | 56 |
| neopentane | 1.5 | 3 | 57 |
| hexachloroethane | 2.56 | 4 | 49 |
| succinonitrile | 1.3 | 5 | 19 |
| cyclohexane | 1.15 | 5 | 57 |

**Table A1.** The value of the entropy of fusion as $\Delta S_{fus}/R$ and the number of rotations (as explained in the text) $n_R$ used to produce Fig. 1. The crystals listed as 'plastic crystals' were identified as such in refs. 17 and 18.

**A2. Data for Fig. 2**

Most of the data for Fig. 2 was taken from ref. 9. The data used to generate the additional point for succinonitrile are presented in Table A2.

| Material | Velocity (m/s) | Relaxation time $\tau_\alpha$ (x $10^{-14}$s) | Lattice spacing a (x $10^{-10}$m) | $\beta \Delta G$ | $\Delta S_{fus}/R$ | $\log k_+^{red}$ |
|---|---|---|---|---|---|---|
| succinonitrile [19] | 5 | 475 | 6.37 | 0.134 | 1.3 | -0.53 |

**Table A2.** The data used to generate the point for succinonitrile in Fig. 2. The relaxation time was calculated from the shear viscosity η using the Maxwell expression, $\tau_\alpha = \eta/E$ where E is the Young's modulus.